\documentstyle[prb,aps]{revtex}
\renewcommand{\ref}[1]{\raisebox{.6ex}{[#1]}}

\newcommand{\be}{\begin{equation}}
\newcommand{\ee}{\end{equation}}
\newcommand{\ba}{\begin{array}}
\newcommand{\ea}{\end{array}}

\begin{document}


\title{ Nernst Effect, Seebeck Effect, and Vortex Dynamics in the Mixed
           State of Superconductors  }

\author{P. Ao   \\ Department of Theoretical Physics                 \\
Ume\aa{\ }University, S-901 87, Ume\aa, SWEDEN  \\ }

\maketitle

\begin{abstract}
Based on the vortex dynamics with the Magnus force and 
a two-fluid model we have derived a set of explicit
expressions between the Nernst and Seebeck coefficients, and the Hall and 
longitudinal resistivities in the linear
response regime of the mixed state of superconductors.
Effects of vortex pinning are included.
The expressions are found to be in agreement with available experimental data.
Present results are valid for large as well as for small Hall angle samples,
and if Hall angle terms are dropped we recover previous theoretical results.
On the other hand, the similar expressions 
based on an alternative vortex dynamics 
with various normal current drag forces on vortices
are in qualitative disagreement with experiments, which implies the  
absence of those normal current drag forces.

\end{abstract}

\noindent
PACS${\#}$s:   74.25.Fy; 74.60.Ec; 74.60.Ge

\section{Introduction }

Up till now, the vortex equation of motion is still a highly 
controversial subject. 
No model seems to be able to give
a consistent explanation of ample experimental data of the related 
transport phenomena of Hall, Nernst, and  Seebeck effects. 
It starts with the obvious observation that 
because the motion of vortices in the mixed state of type II
superconductors generates voltages by the Josephson relation, 
the electric measurements for the longitudinal
and Hall resistivities should be the natural way 
to test vortex dynamics.  
However, for any given effective equation of motion
for vortices, it has been proved to be a difficult task
to calculate the longitudinal resistivity in the mixed state of a 
superconductors, and even more so for the Hall resistivity, 
because of the roles played by vortex pinning and interaction.
It is then not surprising that  no consensus on the basic vortex 
equation of motion has been reached solely based on the tests of electric 
transport measurements.
Thermoelectric transport measurements, the Seebeck and Nernst effects, 
provide additional tests.
A correlation between those two sets of measurements should provide 
a powerful check for the vortex dynamics.
This had been explored before\cite{review}, and
it is worthwhile to point out that there is no report yet
on a direct measurement of the Seebeck effect in a conventional superconductor.
The appearance of the oxide superconductors generates renewed interests in both
vortex dynamics and the thermoelectric measurements.
The experimental data seem to converge\cite{huebener}, 
but theoretical approaches are phenomenological and results are very
diverse\cite{samoilov,wang,dorsey,fischer,ri,meilikhov}. 
Furthermore, there is no proper
account of the pinning effect in the existing theoretical models.

In order to have a better understanding of this problem,
an attempt has been made in the present paper 
to provide a picture to put all those effects into one macroscopic framework, 
based on the vortex dynamics with the Magnus force and a
two-fluid model.
We  will show that this framework accounts well the relevant experiments,
so long as the vortex-antivortex pair fluctions are insignificant.
The core result is a set of explicit relationships
between the electric transport properties, the Hall and longitudinal 
resistivities, and the thermoelectric transport 
properties, the Nernst and Seebeck effects.
Present results are similar to previous ones, 
but important differences remain.
In the course of the comparison between the derived relationships 
and experiments there is no fitting parameter.
Regarding to the  special feature in the thermoelectric transport 
measurements that the magnitudes of supercurrent and normal current are 
equal to each other,
the good agreement between present results and the experimental data strongly
indicates the absence of
normal current drag forces in the vortex dynamic equation.

\section{Derivation Based on the Magnus Force and the Two-fluid Model }

We consider a two-dimensional superconductor film. The magnetic field is 
applied perpendicular to the film. In the presence of temperature gradient
a thermal force will act on a vortex due to the entropy carried by its core.
Quantitatively, for a vortex with a unit length
mass $m_v$ and 
velocity ${\bf v}_l = \dot{\bf r}$, the equation of motion is 
\be
    m_v\dot{\bf v}_l = q_v \frac{n_s(T) h }{2} \; 
      [{\bf v}_{s,total} -{\bf v}_l ]\times \hat{z} - \eta {\bf v}_l 
         + {\bf F}_{pin} + {\bf f} - s_{\phi} \; \nabla T  \; ,
\ee
with vorticity $q_v = \pm 1$, Planck constant $h$,  
the superfluid electron number density $n_s$, 
the vortex friction $\eta$, the pinning force ${\bf F}_{pin}$, 
the fluctuating force ${\bf f}$, and the temperature $T$.
Pinning is due to the inhomogeneity in the sample.
Another inhomogeneous consequence, core-less Josephson vortices in
weak links, will not be considered here, 
although they can be discussed within Eq.(1) with a different
set of transport coefficients.\cite{josephson}
The total superfluid velocity at the vortex is 
${\bf v}_{s,total} ={\bf v}_{s,in} + {\bf v}_{s}$. Here  
${\bf v}_{s,in}$ is the contribution due to other
vortices describing the vortex interaction, 
and ${\bf v}_{s}$ relates to the externally applied supercurrent ${\bf j}_s$ by
\be 
    {\bf j}_s = e \; n_s(T) \; {\bf v}_s \; ,
\ee
with $e$ the electric charge of the carriers.
The existence of the Magnus force, 
the first term at the right hand side of Eq.(1), 
has been shown to be followed from basic properties of a 
superconductor\cite{magnus}. 
The last term at the right hand side of Eq.(1) 
is the thermal force due to the presence of a 
temperature gradient $\nabla T$, with $s_{\phi}$ the unit length entropy
transported by a vortex\cite{review}.
This thermal force may be viewed as  an additional supercurrent  ${\bf j}_T =
e n_s {\bf v}_T $ in Eq.(1), if
\be 
   q_v \frac{n_s(T) h }{2} {\bf v}_T \times 
                   \hat{z} \equiv  - s_{\phi} \; \nabla T \; .
\ee
An alternative vortex equation of motion has been advocated 
recently\cite{kopnin},
with various extra normal fluid dragging force terms. 
We will return to the discussion of its implications
for the Seebeck and Nernst effects later( {\it c.f.} Eqs.(15-22) ).

In the presence a supercurrent vortices will move, and moving vortices
generate the measured electric field by the Josephson relation.
In the steady state and in the linear response regime, 
the average vortex velocity is a linear function of the driven supercurrent.
Because by the Josephson relation the measured electric field ${\bf E}$
is proportional to the average vortex velocity, we have
\be
   {\bf E} = \rho_s \; [{\bf j}_s + {\bf j}_T ] \; ,
\ee
with the superfluid resistivity tensor $\rho_s$ as
\be
   \rho_s = \left( \ba{cc} 
     \rho_{s,xx} & \rho_{s,xy}  \\ 
     \rho_{s,yx} & \rho_{s,yy} 
            \ea \right)  \, .
\ee 
It is should be pointed out that there may be various kinds of vortex motions 
contributing to ${\bf E}$, such as due to vortex vacancies and dislocations. 
Their contributions are additive. 
The presence of the Magnus force may result in a possible large Hall angle 
as well as its sign change.
The absence of this force will rule out the Hall voltage contribution 
of vortices. 
It has been shown that the motion of 
vortex vacancies in a pinned vortex lattice
can lead to the anomalous Hall effect, the sign change of the
Hall angle in the mixed state below the superconducting transition 
temperature, and a consistent explanation for 
the Hall resistivity data may be obtained based on Eq.(1)\cite{anomalous}.
In the rest of the paper no attempt will be made to explicitly calculate 
the superfluid resistivity tensor $\rho_s$. Nevertheless we should pointed
out that  the effects of pinnings are  contained in $\rho_s$.

The presence of the temperature gradient may also generate a normal current 
${\bf j}_n$ because the normal fluid carries entropy. 
By the Ohm's law, we have
\be
   {\bf j}_n = \sigma_n \; [ - S_n \nabla T + {\bf E} ] \; .
\ee
Here $S_n$ is the normal fluid Seebeck coefficient and $\sigma_n$ the normal 
fluid conductivity tensor,
\be
    \sigma_n = \left( \ba{cc} 
     \sigma_{n,xx} & \sigma_{n,xy}  \\ 
     \sigma_{n,yx} & \sigma_{n,yy} 
            \ea \right) \; .
\ee 
A small normal fluid Nernst effect has been neglected here.
In the thermal electric measurements, the total electric current ${\bf j}$
in the sample is zero:
\be
   {\bf j} = {\bf j}_s + {\bf j}_n = 0 \, ,
\ee
with the normal fluid current given in terms of the normal fluid density 
$n_n(T)$
and velocity ${\bf v}_n$ as  ${\bf j}_n = e n_n {\bf v}_n$.
Therefore eliminating both the super and normal currents ${\bf j}_{s,n}$ 
from Eqs.(4,6) and using Eqs.(3,8) we arrive at the following result
\be
   {\bf E} = \rho \; \left[ \sigma_n S_n \nabla T + 
        \frac {2e s_{\phi}}{q_v h} \nabla T\times\hat{z} \right] \; ,
\ee
with the mixed state resistivity tensor $\rho$ as
\be
   \rho = [ \sigma_s + \sigma_n ]^{-1} \; ,
\ee
defined in the electric transport measurements ${\bf E} = \rho \; {\bf j}$ 
with $\nabla T =0$.
Here the superfluid conductivity tensor $\sigma_s$ is given by
$ \sigma_s = \rho_s^{-1} $. 
In an explicit form, we rewrite Eq.(9) as 
\be
   \left( \ba{c} E_x \\ E_y \ea \right) 
     = \left( \ba{cc} 
     S_n [ \rho_{xx} \sigma_{n,xx} + \rho_{xy}\sigma_{n,yx} ]
                  - \frac {2e s_{\phi}}{q_v h} \rho_{xy}   
   & S_n [ \rho_{xx} \sigma_{n,xy} + \rho_{xy}\sigma_{n,yy} ]
               + \frac {2e s_{\phi}}{q_v h} \rho_{xx}  \\
     S_n [ \rho_{yx} \sigma_{n,xx} + \rho_{yy}\sigma_{n,yx} ]
                  - \frac {2e s_{\phi}}{q_v h} \rho_{yy} 
   & S_n [ \rho_{yx} \sigma_{n,xy} + \rho_{yy}\sigma_{n,yy} ]
               + \frac {2e s_{\phi}}{q_v h} \rho_{yx}  \ea \right) 
    \left( \ba{c} \nabla_x T \\ \nabla_y T \ea \right) \, .
\ee
Eq.(11) gives an one-to-one relationship 
between the electric and thermoelectric
measurements in the linear response regime, the core result of the present 
paper. By simultaneous measurements of longitudinal and Hall 
resistivities, and the Seebeck and Nernst effects in one sample,
there will be no adjusting parameter in the
comparison between Eq.(11) and experimental data.
To see this, we note that 
the mixed state resistivity tensor $\rho$ can be obtained by the electric 
measurements, although ideally $ \rho $ could be calculated based on Eq.(1). 
In principle 
the normal fluid Seebeck coefficient $S_n$ and the conductivity tensor 
$\sigma_n$ can be calculated, too\cite{galperin}. 
However, close to the transition temperature both $S_n$ and $\sigma_n$ 
are expect to equal to the corresponding normal state 
values\cite{review}, therefore can be directly measured. 
The only unknown 
quantity now is the unit length entropy $s_{\phi}$ carried by a vortex.
Its value could be calculated by a consideration of the temperature
dependence of the core entropy\cite{maki}.  
In this case the Nernst and Seebeck effects are known from other measurements,
and their measurements will provide a stringent check of the consistency of the
physics leading to Eq.(11).  
Even if $s_{\phi}$ is treated as an unknown quantity, 
the measurement of the Nernst effect can fix its value. Therefore there is no 
fitting parameter for the Seebeck effect. 
We note that 
as the temperature approaches the transition temperature, $\rho \; \sigma_n 
\rightarrow {\bf 1}$, the unit matrix, and $s_{\phi} \rightarrow 0$, we obtain 
the normal state values for the Nernst and Seebeck effects, as expected.

If there is a rotational symmetry in the sample, 
that is, $\rho_{xx} = \rho_{yy}$ and 
 $\rho_{yx} = - \rho_{xy}$,
from Eq.(11), we have a simpler expression for the Seebeck effect as
\be
   \frac{E_x}{\nabla_x T} =  \frac{E_y}{\nabla_y T}  =  
          S_n \; \rho_{xx}\sigma_{n,xx} [ 1 + \tan\theta\tan\theta_n ]
          + \frac{2e s_{\phi} }{q_v h} \; \rho_{xx} \tan\theta  \; ,
\ee
and for the Nernst effect as
\be
    \frac{E_y}{\nabla_x T} =  - \frac{E_x}{\nabla_y T}  = 
     - \frac{2e s_{\phi} }{q_v h} \; \rho_{xx} 
       +   S_n \; \rho_{xx}\sigma_{n,xx} [ \tan\theta - \tan\theta_n ] \; .  
\ee
Here $\theta = \tan^{-1}(\rho_{yx}/\rho_{xx})$ and 
$\theta_n = \tan^{-1}(\rho_{n,yx}/\rho_{n,xx})$ 
are the Hall angles of the mixed state and the normal fluid, respectively.
We should pointed out that the relationship between the vorticity $q_v$, 
the sign of the carrier electric charge $e$, and the direction of the applied 
magnetic field $B$ is 
\be
   q_v \frac{e}{|e|} = \frac{B}{|B|} \; .
\ee
Thus, the Seebeck effect is even with respect to 
the direction of the applied magnetic field, and the Nernst effect is odd,
since the Hall angle is odd for a rotational symmetric sample.
If the differences in the Hall angle terms were ignored, 
expressions similar to Eqs.(12,13)
have been obtained previously by various 
methods\cite{samoilov,wang,dorsey,fischer,ri,meilikhov}.
However, the Hall angle terms can be important, because the Hall angle can be 
in principle close to $\pi/2$ and large Hall angle
data in the mixed state have been indeed reported\cite{fiory,ong}.
The differences between Eqs.(11,12,13) and those obtained in 
Ref.\cite{samoilov,wang,dorsey,fischer,ri,meilikhov} can be therefore 
differentiated experimentally.

\section{Further Discussions }

Although Eq.(11), or the speccial case, Eqs.(12,13), looks simple enough, 
there are quite a few confusions in its derivation and comparison
with experiments. We discuss them below.
  
There are several important differences between Eq.(11) and Eqs.(12,13), 
which are particularly pertinent in the comparison with experiments. 
First, the vortex pinning is automatically included in Eq.(11), as contained
in the mixed state resistivity tensor $\bf \rho$. 
Secondly, as a special case of the first point, 
effects of the vortex guided motion due to a possible correlated pinnings 
is included. Therefore Eq.(11) also
applies to the case of no rotational symmetry in a sample.   
The guided motion not only generates a large even part with respect 
to the direction of the applied magnetic field in $\rho_{yx,xy}$, also
may result in $\rho_{xx} \neq \rho_{yy}$. 
A special attention should be paid to this effect in an interpretation of 
experimental data.
Thirdly, another special case of the first point, Eq.(11)
applies to the case of the anomalous Hall effect, because with 
considerations of the vortex many-body correlation and pinning effect
Eq.(1) can lead to this effect\cite{anomalous}.
Finally, it should be emphasized that the base for Eq.(11) is Eq.(1), and 
an alternative vortex dynamics model will generally give a difference 
expression.({\it c.f.} Eq.(18)).
We also note that the present
results are concrete in contrast to those based on general symmetry 
considerations.\cite{symmetry}

Two remarks on 
different kinds of vortex motions are in order.
First, in the derivation of Eq.(11) we notice that the effective supercurrent
${\bf j}_T$ in Eq.(3) due to the thermal force should 
carry the sign of the vorticity, but 
it has been treated as if independent of it as the real supercurrent 
${\bf j}_s$.
Although different vortex motions 
have different contributions to the superfluid
resistivity tensor $\rho_s$, as long as the motions are all
vortex like, there is no need to concern with the vorticity dependence 
in ${\bf j}_T$.
Interestingly, this is also true for the vacancy motion,
because of the absence of a normal core 
for a vacancy as pointed out in Ref.\cite{anomalous} in the 
discussion of the anomalous Hall effect.  
However, it is not true for antivortices, and leads to the second remark.
We note that the thermal force, Eq.(3), 
is the same for both vortices and antivortices. 
Therefore the effective supercurrent ${\bf j}_T$ takes opposite signs 
for vortices and antivortices, respectively.
Then in the connection between the electric and the 
thermoelectric transport measurements, the contributions due to vortices and 
antivortices should be carefully separated out, as noted by 
Ri {\it et al.}\cite{huebener}. 
Fortunately, the generating of 
free antivortices through vortex pair fluctuations 
is energetically the most unfavorable one
below the transition temperature. Therefore it can be ignored below the
transition temperature (Kosterlitz-Thouless transition in 2-d cases).

It is appropriate to further discuss 
the important differences between the present results and previous work.
One of them has been stated above: the proper consideration of the
Hall effect terms. Because of this, it appears that 
present results, Eq.(11) or Eqs.(12,13), are different from all those in
previous work.
For example, in Ref.\cite{ri} it has been suggested that $\tan\theta =
\tan\theta_n$, in Ref.\cite{dorsey,fischer} 
there is no second term in Eq.(13), and in Ref.\cite{samoilov} the 
contribution of normal fluid resistivity is not present.
The approach of Ref.\cite{samoilov} is inappropriate regarding to 
the use of the
Ohm's law for the normal fluid: it fails to consider the electric field,
and the mixed state resistivity used there is actually given by
superfluid resistivity tensor $\rho_s$,
not by the two-fluid type $\rho$ as given by the present Eq.(10).
The pinning has not been discussed in Refs.\cite{samoilov,dorsey,fischer,ri}.
In Ref.\cite{wang}, 
the pinning effect has been considered, but it is found there that
there is a drastic change of equation of motion for a vortex, and,
in their Bardeen-Stephen limit a violation of the Onsager relation
has been found.
The approach in Ref.\cite{meilikhov} 
is very similar to that of the present paper.
However, the effect of the pinning has been assumed to be modeled by 
percolation. Further, three additional parameters, $N_s/N_n, f_s, f_n$,
have been introduced into their model,
which, in the light of the present model, looks redundant.

Now we discuss the comparison of Eq.(11) with  experiments 
with an emphasis on features overlooked before.
It is a general experimental observation 
that Eqs.(12,13) or the similar ones in the literature are consistent with
experimental data.\cite{review,huebener}. 
A careful analysis has revealed a small discrepancy, $ \sim 20 \% $
depending on applied magnetic fields and samples,
in the comparison of Eq.(12) with the Seebeck effect data of the
oxide superconductor YBCO samples, peaked
at a temperature about 10K below the transition temperature\cite{ri}. 
There are many possible mechanisms responsible for this discrepancy, 
such as the decrease of $S_n$ as lowering the temperature 
calculated in Ref.\cite{galperin}.
In view of the above discussion of the guided motion, 
its effects are also likely responsible for this discrepancy.
For the Hall angle term in Eq.(13), a careful measurement 
reveals that it has no observable effect near the transition 
temperature\cite{ri}: The measured normal Hall angle is 
$\tan\theta_n \sim 0.01$, but the Nernst effect 
data require that $|\tan\theta - \tan\theta_n| << 0.005$.  
This is bored out by Eqs.(11,13): 
As the temperature approaches the transition temperature, 
$\rho \rightarrow \rho_n$, therefore $\tan\theta \rightarrow 
\tan\theta_n $, the Hall angle contribution to the Nernst effect then
disappears.
It should be pointed out that this disapparence of Hall angle terms 
in Eq.(13) occurs 
just near the transition temperature, not in the whole mixed state regime
as claimed in Ref.\cite{ri}.
It is also interesting to pointed out that this suggests $\sigma_s = 0$
in the mixed state as approaching the transition temperature from below.
Based on the above discussions we concluded that Eq.(11) is in good agreement 
with available experimental data.

We note that in the above discussion
only the macroscopic anisotropy due to the vortex 
guide motion has been considered.
This may not be enough, because of the microscopic anisotropy due to
the electronic and lattice structures as well as the symmetry of the
pairing wavefunction. For such cases, both $s_{\phi}$ and $S_n$ 
should be represented by tensors. 
The vortex viscosity $\eta$ as well as the Magnus force may also adopt
tensor forms. The formal equation, Eq.(9), however, will remain unchanged, but
the detailed equation, Eq.(11), will become quite complicated.
We will no pursue this question any further, 
but simply mention here that this may
have been indicated experimentally in an oxide superconductor\cite{yu}. 

One point we should emphasize here is that, 
although the Hall angle is usually small, 
large Hall angles in the mixed state, 
$|\tan\theta| \sim 1$, have been reported for both conventional\cite{fiory} 
and oxide\cite{ong} superconductors. Simultaneous measurements on such samples
will provide a unique test of Eq.(11). Further experiments are clearly 
needed in this direction.

\section{Alternative Vortex Dynamics Model }

As mentioned at the beginning of the paper and expressed 
by Eq.(8), the magnitudes of super and normal currents are equal to each other.
This is quite different from the electric measurements where the normal
current is usually much smaller than the supercurrent. 
This feature will allow us to study rather conveniently
the consequence of a possible normal current drag force.
If there were a drag force due to a normal current in Eq.(1),
its effect can be effectively 
treated as an additional supercurrent in Eq.(1), as for the case of
the thermal force in Eq.(3). 
Then we would simply repeat the calculation leading to
Eq.(11), or Eqs.(12,13), and obtain modified ones. 
Therefore the effects of the normal current drag force would 
appear in the modified equations, which  can be compared
with experiments. We present such an analysis below.

A recently proposed alternative vortex dynamics model 
with various normal fluid dragging force terms
has the following form\cite{kopnin}:
\be
   0 = q_v \frac{n_s(T) h }{2} \; 
      [{\bf v}_{s,total} -{\bf v}_l ]    \times \hat{z} 
      - D' \; [{\bf v}_{n} -{\bf v}_l ]\times \hat{z} 
      + D  \; [{\bf v}_{n} -{\bf v}_l ] \, .
\ee
For usual superconductors, the numerical parameter $\omega_0 \tau << 1$. 
In this case it has been found that 
$ D' = \frac{q_v h }{2} [ n_e - n_n ]$ and 
$D = \omega_0 \tau \frac{ h }{2} n_s$, with $n_e = n_n + n_s$ the total
fluid density.\cite{kopnin}
Putting the pinning force, the noise, and the thermal driving force into
 Eq.(15),
and expressing the super and normal fluid velocities by the corresponding
electric currents,
the equivalent alternative vortex dynamics equation is:
\[
  0 = - \alpha  q_v \frac{n_s(T) h }{2} \; {\bf v}_l \times \hat{z} 
      - D  {\bf v}_l + q_v \frac{n_s(T) h }{2}\;{\bf v}_{s,in}\times \hat{z} 
      + ( {\bf j}_s + {\bf j}_n ) \times {\bf \Phi}_0  
      -  \frac{n_e}{n_n} {\bf j}_n  \times {\bf \Phi}_0  
\]
\be
      + \omega_0 \tau \frac{n_s}{n_n}
    \left( \frac{\bf B } {|B| } \times {\bf j}_n \right)\times {\bf \Phi}_0  
      + {\bf F}_{pin} + {\bf f} - s_{\phi} \; \nabla T  \, .
\ee
Here ${\bf \Phi}_0$ is the magnetic flux quantum in the direction of
the magnetic field ${\bf B}$, and $\alpha$ a numerical parameter whose
precise value is unimportant presently.
For this alternative vortex dynamics model, 
the measured electric field in the linear response regime is 
\be
   {\bf E} = \rho^a_s \left[ {\bf j}_s + {\bf j}_n -  
             \frac{n_e}{n_n} \; {\bf j}_n 
      + \omega_0 \tau \frac{n_s}{n_n} 
        \frac{\bf B}{|B|} \times {\bf j}_n \right] \, .
\ee
Here $\rho^a_s $ may be called the superfluid resistivity tensor, and
the superscript $a$ standard for the alternative vortex dynamics.
Eliminating the super and normal currents with the aid of the 
Ohm's law, Eq.(6), and of the two fluid model, Eq.(8), 
we have,
\be
   {\bf E} = \rho^a \; \left[ \frac{n_e}{n_n} \sigma_n S_n \nabla T + 
        \frac {2e s_{\phi}}{q_v h} \nabla T\times\hat{z} 
        - \omega_0 \tau \frac{n_s}{n_n}
            \frac{ { \bf B} }{|B| } \times 
             \sigma_n S_n \nabla T \right] \; ,
\ee
with the mixed state resistivity tensor
\be
    \rho^a = \left[ (\rho_s^a)^{-1} + \frac{n_e}{n_n}\sigma_n 
                 - \omega_0 \tau \frac{n_s}{n_n}
            \frac{ { \bf B} }{|B| } \times 
             \sigma_n \right]^{-1} \, .
\ee
We note that here Eq.(19) is completely different from 
the usual two fluid resistivity formula of Eq.(10).   

Assuming the rotational symmetry in the sample, the expressions similar to 
Eqs.(12,13) for the alternative vortex dynamics are
\be
   \frac{E_x}{\nabla_x T}  =  
          S_n \; \rho_{xx}^a \sigma_{n,xx} \left[ 
           \frac{n_e}{n_n} ( 1 + \tan\theta^a \tan\theta_n) 
           + \omega_0 \tau \frac{n_s}{n_n}
     \frac{B}{|B| }(\tan\theta^a - \tan\theta_n) \right]
          + \frac{2e s_{\phi} }{q_v h} \; \rho_{xx}^a \tan\theta^a \; ,
\ee
and for the Nernst effect as
\be
    \frac{E_y}{\nabla_x T}  = 
     - \frac{2e s_{\phi} }{q_v h} \; \rho_{xx}^a  
       +   S_n \; \rho_{xx}^a \sigma_{n,xx} \left[ 
         \frac{n_e}{n_n} ( \tan\theta^a - \tan\theta_n) 
       -  \omega_0 \tau \frac{n_s}{n_n}
     \frac{B }{ |B| } ( 1 + \tan\theta^a \tan\theta_n)  \right] \; .  
\ee
For the case of small Hall angle terms, where $\omega_0 \tau << 1$, 
we have the ratio 
\be
  \tan\theta_{th}^a
  = \frac{ \frac{E_y}{\nabla_x T} }{  \frac{E_x}{\nabla_x T}  } 
  = - \frac{n_n}{n_e}  \frac{ 2e s_{\phi} }{q_v h}  
        \frac{ \rho_{n,xx} }{ S_n } \, .
\ee
The same ratio from Eqs.(12,13) is
\be
  \tan\theta_{th}
  = - \frac{ 2e s_{\phi} }{q_v h}  
        \frac{ \rho_{n,xx} }{ S_n } \, .
\ee

In the low temperature limit, $\frac{n_n}{n_e} \rightarrow 0$.
The normal fluid Seebeck coefficient also goes to zero in the same way as 
$n_n$.\cite{galperin} 
The entropy $s_{\phi}$ carried by a vortex, $s_{\phi} \rightarrow 0$  
in a rate slower than that of  $n_n$, 
because its rate is determined by the core
level spacing, not by the energy gap as in $n_n$.
Therefore, as lowering the temperature to zero, 
the ratio $\tan\theta_{th}^a$ given by Eq.(22) goes to zero, and the ratio 
$\tan\theta_{th}$ given by Eq.(23) goes to infinite.
Experimentally, it has been found that the magnitude of the ratio 
goes to infinite in the low temperature limit.\cite{putti}
This is in agreement with Eq.(23).
This also shows that Eq.(22) is in qualitative disagreement with experimental 
observations.
Since Eq.(22) is a direct consequence of Eq.(15), one concludes 
that this disagreement implies 
the alternative vortex dynamics model, Eq.(15), is inconsistent with
experiments.

\section{Summary }

To summarize, based on the vortex dynamic equation with the Magnus force and 
a two-fluid model,
relationships between Nernst, Seebeck effects, and the Hall
and longitudinal resistivities have been derived in the linear response regime.
A unified macroscopic description of the
electric and thermoelectric coefficients in the mixed state of superconductors
has been obtained. 
When the Hall angle terms become negligible, we recover previous results. 
The present derived relationships is consistent with existing experiments. 
Therefore it provides an additional experimental confirmation for the 
underlining vortex dynamic equation.
The alternative vortex dynamical model with various
extra normal fluid dragging force terms 
is in qualitative disagreement with experiments.

\noindent
{\bf Acknowledgments: } 
The author appreciates stimulating discussions with Andrei Shelankov.
Part of the work was done during a visit to 
the US National High Magnetic Field Lab at Tallahassee. 
The hospitality of NHMFL as well as
a travel aid from Swedish Natural Science Research Council 
are gratefully acknowledged.

\end{document}